\documentclass[12pt]{article}
\usepackage{graphicx}
\def \b{{\cal B}}
\def \bea{\begin{eqnarray}}
\def \beq{\begin{equation}}
\def \eea{\end{eqnarray}}
\def \eeq{\end{equation}}
\def \ko{K^0}
\def \ok{\overline{K}^0}
\def \s{\sqrt{2}}
\def \st{\sqrt{3}}
\def \sx{\sqrt{6}}

\begin{document}
\rightline{EFI-06-13}
\rightline{hep-ph/0607346}
\rightline{July 2006}
\bigskip
\centerline{\bf INTERFERENCE BETWEEN DOUBLY-CABIBBO-SUPPRESSED AND}
\smallskip
\centerline{\bf CABIBBO-FAVORED AMPLITUDES IN $D^0 \to K_S (\pi^0,\eta,\eta')$
DECAYS}
\bigskip

\centerline{Jonathan L. Rosner\footnote{rosner@hep.uchicago.edu}}
\centerline{\it Enrico Fermi Institute and Department of Physics}
\centerline{\it University of Chicago, 5640 S. Ellis Avenue, Chicago, IL 60637}

\begin{quote}
A definite relative phase and amplitude exists between the doubly-Cabibbo-%
suppressed amplitude for $D^0 \to \ko M^0$ and the Cabibbo-favored amplitude
for $D^0 \to \ok M^0$, where $M^0 = (\pi^0,\eta,\eta')$: $A(D^0 \to \ko M^0) =
- \tan^2 \theta_C A(D^0 \to \ok M^0)$.  Here $\theta_C$ is the Cabibbo angle.
This relation, although previously recognized (for $M^0 = \pi^0$) as a
consequence of the U-spin subgroup of SU(3), is argued to be less sensitive to
corrections involving SU(3) breaking than related U-spin relations involving
charged kaons or strange $D$ mesons.
A corresponding relation between $D^+ \to \ko \pi^+$ and $D^+ \to \ok \pi^+$ is
not predicted by U-spin.  As a consequence, one expects the asymmetry
parameters $R(D^0,M^0) \equiv [\Gamma(D^0 \to K_S M^0) - \Gamma(D^0 \to K_L
M^0)/[\Gamma(D^0 \to K_S M^0) + \Gamma(D^0 \to K_L M^0)]$ each to be equal to
$2 \tan^2 \theta_C = 0.106$, in accord with a recent CLEO measurement $R(D^0)
\equiv R(D^0,\pi^0) = 0.122 \pm 0.024 \pm 0.030$.  No prediction for the
corresponding ratio $R(D^+)$ is possible on the basis of U-spin.
\end{quote}

\leftline{PACS numbers:  13.25.Ft, 11.30.Hv, 14.40.Lb}
\bigskip

The large number of flavor-tagged neutral $D$ mesons collected by the CLEO
Collaboration has permitted unprecedented studies of branching fractions,
shedding light on details of the Cabibbo-Kobayashi-Maskawa matrix, flavor
mixing, and signatures for new physics.  Recently these data have been
analyzed for the decays $D \to K_S \pi$ and $D \to K_L \pi$ \cite{He:2006gd}.
Whereas the rate asymmetry
\beq
R(D^0) \equiv \frac{\Gamma(D^0 \to K_S \pi^0) - \Gamma(D^0 \to K_L \pi^0)}
 {\Gamma(D^0 \to K_S \pi^0) + \Gamma(D^0 \to K_L \pi^0)}
\eeq
is found to be non-zero, $R(D^0) = 0.122 \pm 0.024 \pm 0.030$, the
corresponding asymmetry for $D^+$ decays,
\beq
R(D^+) \equiv \frac{\Gamma(D^+ \to K_S \pi^+) - \Gamma(D^+ \to K_L \pi^+)}
 {\Gamma(D^+ \to K_S \pi^+) + \Gamma(D^+ \to K_L \pi^+)}
\eeq
is consistent with zero:  $R(D^+) = 0.030 \pm 0.023 \pm 0.025$.  In this note
I shall show that one expects on general grounds a definite value $R(D^0) =
2 \tan^2 \theta_C \simeq 0.106$, where $\theta_C$ is the Cabibbo angle:
$\tan \theta_C \simeq 0.230$, while in general no such prediction is possible
for $R(D^+)$.  Moreover, $R(D^0,\eta) = R(D^0,\eta') = 2 \tan^2 \theta_C$ is
predicted independently of the flavor-octet/flavor-singlet makeup of $\eta$
and $\eta'$.  This picture remains valid for a more general representation of
$\eta$ and $\eta'$ involving flavor-symmetry breaking \cite{Feldmann:1998vh}.

The possibility of interference between Cabibbo-favored (CF) decays of charmed
mesons to $\ok + X$ and doubly-Cabibbo-suppressed (DCS) decays to $\ko + X$ was
noted in Refs.\ \cite{Buccella:1992ym} and \cite{Bigi:1994aw}.  For the decays
$D \to K_{S,L} \pi$ asymmetries $R(D^{0,+}) \simeq 2 \tan^2 \theta_C$ were
anticipated \cite{Bigi:1994aw}, with the relation expected to be more exact for
$D^0$.  We shall show that $R(D^0) = 2 \tan^2 \theta_C$ is predicted by the
U-spin \cite{Usp} subgroup of SU(3) \cite{Wolfenstein:1995kv,Lipkin:1998in,%
Gronau:2000ru} without identifiable SU(3)-violating corrections, whereas a
corresponding relation for $R(D^+)$ is not predicted by U-spin.

The U-spin argument \cite{Gronau:2000ru} proceeds as follows.  The initial
$D^0 = c \bar u$ state is a U-spin singlet because it contains no $d$ or $s$
quarks.  The Cabibbo-favored transition $c \to s u \bar d$ has $\Delta U
= - \Delta U_3 = 1$ while the doubly-Cabibbo-suppressed $c \to d u \bar s$
transition, with amplitude $-\tan^2 \theta_C$ relative to the first, has
$\Delta U = \Delta U_3 = 1$.  Thus, the two transitions lead to $U=1$ final
states which are U-spin reflections of one another.

Now consider the final states consisting of $\ok M^0$ or $\ko M^0$, where
$M^0 = (\pi^0,\eta,\eta')$, with
\beq \label{eqn:mix}
\eta = \eta_8 \cos \theta + \eta_1 \sin \theta~,~~
\eta' = - \eta_8 \sin \theta + \eta_1 \cos \theta~~;
\eeq
\beq
\eta_8 \equiv \frac{1}{\sx} \left( 2 s \bar s - u \bar u - d \bar d \right)~,~~
\eta_1 \equiv \frac{1}{\st} \left( s \bar s + u \bar u + d \bar d \right)~~.
\eeq
A reasonable representation of octet-singlet mixing in $\eta$ and $\eta'$
is obtained for $\sin \theta \simeq - 1/3$ \cite{Kawarabayashi:1980dp,%
Gilman:1987ax,Chau:1990ay,Dighe:1995gq} but our results will be not only
independent of $\theta$ but valid even for a more general picture of $\eta$
and $\eta'$ than Eq.\ (\ref{eqn:mix}) \cite{Feldmann:1998vh}.

The $\pi^0$ and $\eta_8$ are admixtures of U-spin
singlets and triplets with $U_3 = 0$.  Because of Bose symmetry, the U-spin
triplets, when combined with final-state neutral kaons which necessarily have
$U=1$ and $U_3 = \pm 1$, can only form states of total $U=2$, which are not
produced in the $c \to s u \bar d$ or $c \to d u \bar s$ transitions.
Consequently, only the U-spin singlet projections of $\pi^0$ and $\eta_8$
contribute to the decays $D^0 \to \ok M^0$ and $D^0 \to \ko M^0$.

The flavor-singlet component $\eta_1$, when combined with the neutral kaon,
necessarily gives a state with $U=1$.  Thus any state $\ko M^0$ or $\ok M^0$
produced in $D^0$ decay, with $M^0 = (\pi^0,\eta,\eta')$, is a state with
$U=1$ and $U_3 = \pm 1$.  As a result, symmetry under U-spin reflection implies
\beq \label{eqn:U0}
\frac{A(D^0 \to \ko M^0)}{A(D^0 \to \ok M^0)} = - \tan^2 \theta_C~~~.
\eeq
This result does not depend upon any specific picture of $\eta$--$\eta'$
mixing but only on U-spin.  It remains valid even when Eq.\ (\ref{eqn:mix})
is replaced by a more general representation of $\eta$ and $\eta'$ based
on two mixing angles rather than one, required in a consistent 
treatment of flavor symmetry breaking \cite{Feldmann:1998vh}.

\begin{figure}
\mbox{\includegraphics[width=0.47\textwidth]{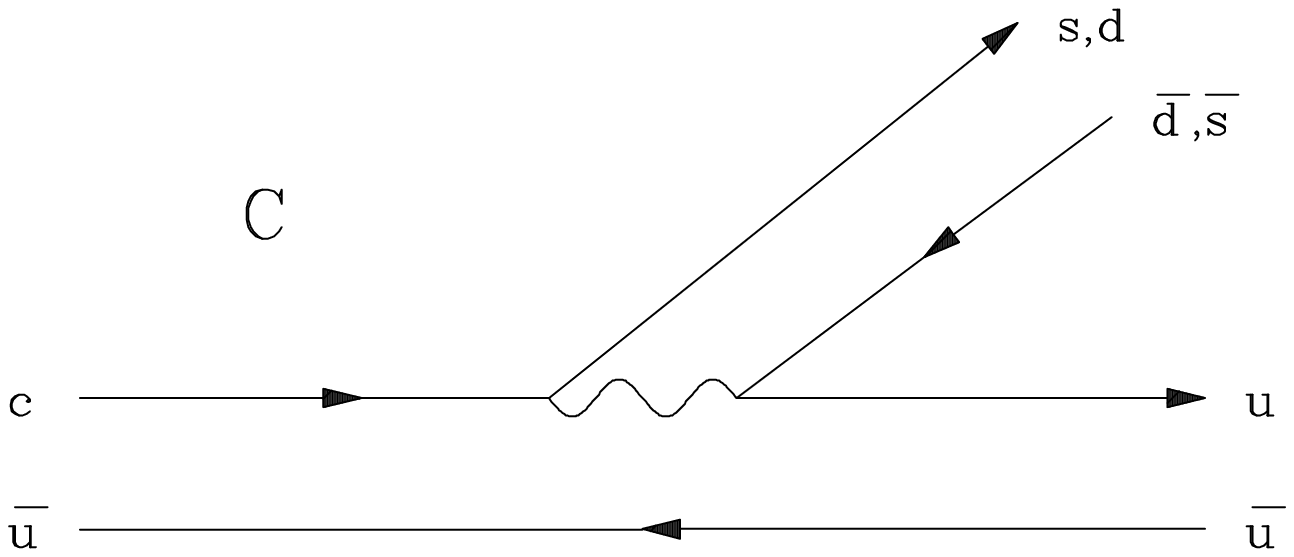} \hskip 0.2in
\includegraphics[width=0.47\textwidth]{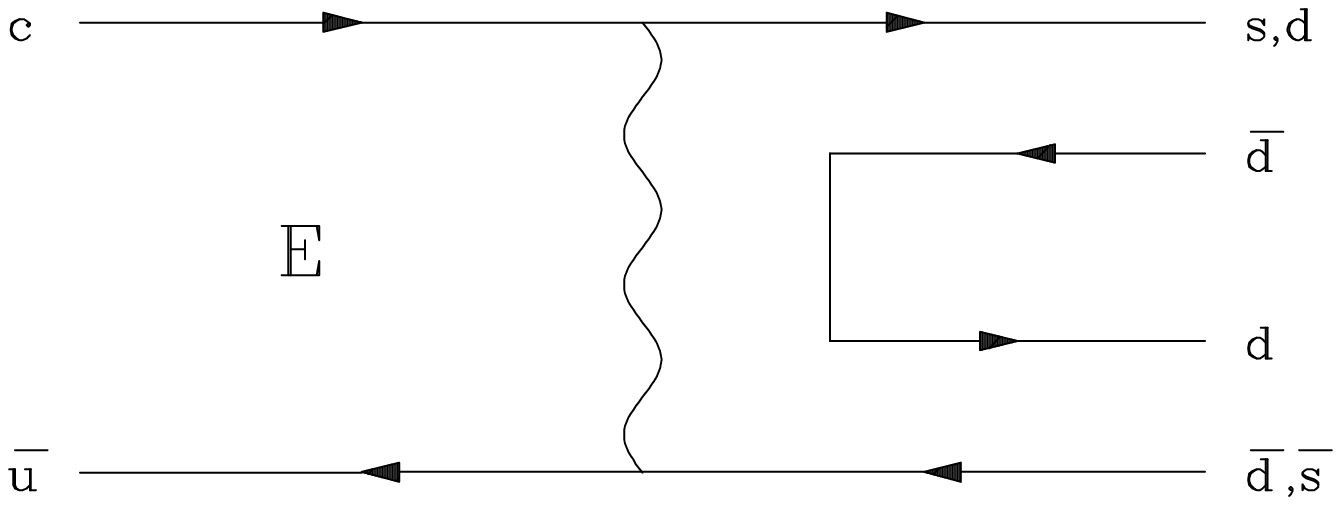}}
\caption{Diagrams contributing to $D^0 \to \ok M^0$ and $D^0 \to \ko M^0$.
Left:  color-suppressed ($C$); right: exchange ($E$).
\label{fig:CandE}} 
\end{figure}

Eq.\ (\ref{eqn:U0}) does not appear to receive any corrections associated
with flavor-SU(3) breaking.  In the language of flavor diagrams
\cite{Rosner:1999xd,Chiang:2001av,Chiang:2002mr}, the amplitudes for $D^0 \to
\ok M^0$ and $D^0 \to \ko M^0$ are both linear combinations of the reduced
amplitudes $C$ and $E$, differing by an overall factor of $-\tan^2 \theta_C$.
$C$ is a color-suppressed amplitude in which the subprocess $c \to s u \bar d$
or $c \to d u \bar s$ is followed by the incorporation of the $s \bar d$ into a
$\ok$ or the $d \bar s$ into a $\ko$.  These processes are expected to occur
with equal amplitude and phase.  $E$ is an exchange amplitude involving the
spectator $\bar u$ quark in the $D^0$ in the subprocess $c \bar u \to s \bar d$
(Cabibbo-favored) or $c \bar u \to d \bar s$ (doubly-Cabibbo-suppressed).
These diagrams are illustrated in Fig.\ \ref{fig:CandE}.  Assuming that the
four-fermion interaction mediated by the $W$ (depicted by a wiggly line) is
local, the evolution of the $s \bar d$ system into $\ok M^0$ should be
characterized by the same amplitude and strong phase as that of $d \bar s$ into
$\ko M^0$.  There may be short-distance flavor-dependent QCD corrections to the
four-fermion interactions, but we cannot identify any important long-distance
sources of SU(3) breaking in the U-spin relation (\ref{eqn:U0}).

\begin{figure}
\includegraphics[width=0.8\textwidth]{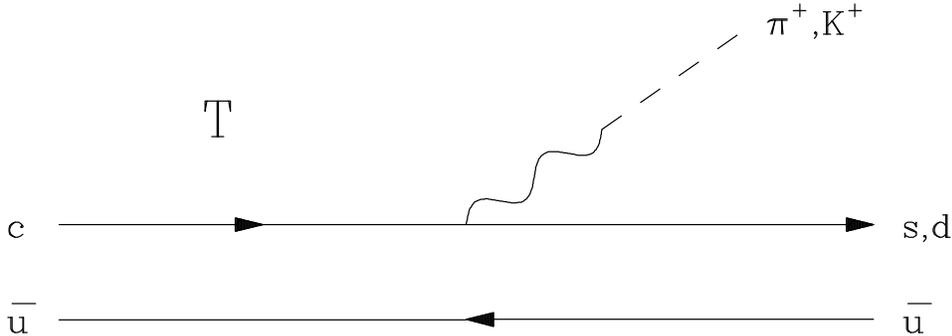}
\caption{Color-favored diagram contributing to $D^0 \to K^- \pi^+$ and $D^0 \to
K^+ \pi^-$.
\label{fig:T}}
\end{figure}

Other U-spin relations noted in Ref.\ \cite{Gronau:2000ru}, namely
\beq \label{eqn:Uc}
\frac{A(D^0 \to K^+ \pi^-)}{A(D^0 \to K^- \pi^+)} =
\frac{A(D^+ \to \ko \pi^+)}{A(D_s^+ \to \ok K^+)} =
\frac{A(D_s^+ \to \ko K^+)}{A(D^+ \to \ok \pi^+)} = - \tan^2 \theta_C~~~.
\eeq
do not appear immune to SU(3) breaking.  The second and third involve spectator
quarks with different masses and thus one expects them to be characterized by
different form factors.  The first involves amplitudes of the form $T+E$,
where $E$ is an exchange amplitude as noted above and $T$ is a color-favored
``tree'' (or factorized) amplitude involving the subprocess $c \to \pi^+ s$
(Cabibbo-favored) or $c \to K^+ d$ (doubly-Cabibbo-suppressed) as depicted in
Fig.\ \ref{fig:T}.  The ratio in the first term of Eq.\ (\ref{eqn:Uc}) thus
involves ratios of decay constants $f_K/f_\pi$ and form factors $F(D \to
\pi)/F(D \to K)$ each of which can differ substantially from unity.  (See
the remarks in Ref.\ \cite{Lipkin:1998in}.) The observed ratio \cite{He:2006gd}
$r^2_{K \pi} \equiv \b(D^0 \to K^+ \pi^-)/\b(D^0 \to K^- \pi^+)$ is $0.00363
\pm 0.00038$, about $2.2 \sigma$ above its value of $\tan^4 \theta_C = 0.00279$
predicted by U-spin.  Rescattering processes $K^- \pi^+ \to \ok M^0$ and $K^+
\pi^- \to \ko M^0$ can lead to contributions topologically equivalent to the
$E$ diagram.  These processes, if important, could lead to some violation of
the U-spin relation between $D^0 \to \ko M^0$ and $D^0 \to \ok M^0$.

The amplitudes for $D^+ \to \ko \pi^+$ and $D^+ \to \ok \pi^+$ are related upon
U-spin reflection to amplitudes for $D_s^+ \to \ok K^+$ and $D_s^+ \to \ko
K^+$, respectively, and not to one another.  They do not have the same
flavor-SU(3) decomposition.  One finds instead \cite{Chiang:2001av,%
Chiang:2002mr} $A(D^+ \to \ko \pi^+) = C + A$ while $A(D^+ \to \ok \pi^+)=T +
C$ aside from an overall ratio $-\tan^2 \theta_C$.  Here $A$ is an annihilation
amplitude involving the spectator quark.  The process $c \bar d \to u \bar s$
is followed by the evolution of the $u \bar s$ pair into $\ko \pi^+$.  Thus
without further flavor-SU(3) analysis (for example, by updating the results
of \cite{Chiang:2001av,Chiang:2002mr}) it is impossible to predict the
amplitude ratio $A(D^+ \to \ko \pi^+)/A(D^+ \to \ok \pi^+)$.

The phase conventions in which the above amplitudes have been expressed are
such that the CP eigenstates of neutral kaons (neglecting CP violation) are
\cite{Bigi:1994aw}
\beq
K_S = \frac{1}{\s}\left( \ok - \ko \right)~~,~~~
K_L = \frac{1}{\s}\left( \ok + \ko \right)~~~.
\eeq
The $\ok$ and $\ko$ contributions are thus, according to Eq.\ (\ref{eqn:U0}),
expected to add constructively for $D^0 \to K_S M^0$ and destructively for
$D^0 \to K_L M^0$, leading (in first order of the ratio of DCS to CF
amplitudes) to
\beq \label{eqn:asym}
R(D^0,M^0) =  2 \tan^2 \theta_C \simeq 0.106
\eeq
as noted.  This relation should hold not only for $M^0 = \pi^0$ but also for
$M^0 = (\eta,\eta')$, independently of the makeup of $\eta$ and $\eta'$ and
of any flavor symmetry violation in their description.
\bigskip

I thank D. Asner, Y. Grossman and H. Lipkin for helpful comments.  This work
was supported in part by
the United States Department of Energy through Grant No.\ DE-FG02-90ER-40560.

\end{document}